\documentclass[mathleft
]{an}
\usepackage{graphicx}
\usepackage{times}
\usepackage{bm}
\begin{document}
\sloppy
\setlength{\mathindent}{0pt}
\newcommand{\ve}[1]{\boldsymbol{#1}}
\newcommand{\kef}{k_{\rm f}}
\newcommand{\etat}{\eta_{\rm t}}
\newcommand{\urms}{u_{\rm rms}}
\newcommand{\Omm}{\Omega_{\rm m}}
\newcommand{\brac}[1]{\langle #1 \rangle}
\newcommand{\mean}[1]{\overline{#1}}

\def\onethird{{\textstyle{1\over3}}}
\def\onehalf{{\textstyle{1\over2}}}

% The following seven commands are intended for editorial usage and should be ignored by
% the author(s).
\Pagespan{34}{}% Document's page range. 
% If second parameter is left empty, the last page is computed automatically.
\Yearpublication{2010}%
\Yearsubmission{2009}%
\Month{9}%   
\Volume{331}%  
\Issue{1}% 
 \DOI{10.1002/asna.200911254}% 

\title{Influence of Ohmic diffusion on the excitation and dynamics of MRI}

\author{M.J. Korpi\inst{1}\fnmsep\thanks{Corresponding author:
Maarit.Korpi@helsinki.fi} \and P.J. K\"apyl\"a\inst{1,2} \and M.S.
V\"ais\"al\"a\inst{1}}
%Example 
%for footnote, note the usage of the \texttt{fnmsep}
%command as separator between institute number and footnote mark} 
%\and M. J. Korpi$^1$
%\and P. J. K\"apyl\"a$^1$
%\and A. Brandenburg$^2$
%}
\titlerunning{Influence of Ohmic diffusion on the excitation and dynamics of MRI}
\authorrunning{M.J. Korpi, P.J. K\"apyl\"a \& M.S. V\"ais\"al\"a }
\institute{
Observatory, University of Helsinki, PO BOX 14, FI-00014 University of Helsinki, Finland
\and 
NORDITA, AlbaNova University Center, Roslagstullsbacken
              23, SE-10691 Stockholm, Sweden}

\received{2009 Sep 9}
\accepted{2009}
\publonline{2009 Dec 30}


\abstract{%
  In this paper we make an effort to understand the interaction of
  turbulence generated by the magnetorotational instability (MRI) with
  turbulence from other sources, such as supernova explosions (SNe) in
  galactic disks. First we perform a linear stability analysis (LSA)
  of non-ideal MRI to derive the limiting value of Ohmic diffusion
  that is needed to inhibit the growth of the instability for
  different types of rotation laws. With the help of a simple analytical
  expression derived under first-order smoothing approximation (FOSA), 
  an estimate of the limiting turbulence level and hence the turbulent 
  diffusion needed to damp the MRI is derived. Secondly, we perform
  numerical simulations in local cubes of isothermal nonstratified gas
  with external forcing of varying strength to see whether the linear
  result holds for more complex systems. 
  Purely hydrodynamic calculations with forcing, rotation and shear
  are made for reference purposes, and as expected, non-zero Reynolds
  stresses are found. In the magnetohydrodynamic calculations,
  therefore, the total stresses generated are a sum of the
  forcing and MRI contributions. 
  To separate these contributions, we perform reference runs with MRI-stable
  shear profiles (angular velocity increasing outwards),
  which suggest that the MRI-generated stresses indeed
  become strongly suppressed as function of the forcing. 
  The Maxwell to Reynolds stress ratio is observed to decrease by an
  order of magnitude as the turbulence level due to external forcing
  exceeds the predicted limiting value, which we interpret as a
  sign of MRI suppression.
  Finally, we apply these
  results to estimate the limiting radius inside of which the SN
  activity can suppress the MRI, arriving at a value of 14 kpc.}

\maketitle

\section{Introduction}

  Most differentially rotating astrophysical disks are
  susceptible to the magnetorotational instability (MRI). In accretion
  disks this instability is likely to be the most
  important source of turbulence and angular momentum transport
  leading to mass accretion (originally proposed by Balbus \& Hawley
  1991). It has also been proposed that the MRI would play a role in
  driving turbulence in the outer parts of galactic disks where the
  stellar forcing is of minor importance (Sellwood \& Balbus 1999; see
  also Piontek \& Ostriker 2007 and references therein).

  Whether the powerful stellar forcing suppresses or interacts with
  the galactic MRI in the inner parts of galactic disks is still an open
  question. The study by Workman \& Armitage (2008), who investigated
  the interaction of the MRI with nonhelically forced turbulence in
  isothermal nonstratified local cubes of gas with Keplerian
  differential rotation, can be regarded as a first step to understand
  this issue. They found that external turbulence can indeed
  modify the turbulent transport when the level of turbulence becomes
  greater than the strength of the MRI-generated turbulence, and that the
  degree of the influence is dependent on the forcing scale. Only
  turbulence forced at large wavenumbers (smallest scales) was
  interpreted to suppress MRI, while large-scale forcing seemed to
  enhance it. In this study we show that the forcing wavenumber
  dependence can be understood with the help of the linear stability
  analysis. We also show that reference calculations with
  MRI-stable rotation profiles are needed to fully determine the level
  of turbulence and angular momentum transport that results from the MRI
  in comparison to other sources of turbulence.

  The paper is organised as follows. In Sect.~\ref{LSA} we present a
  linear stability analysis of non-ideal MRI to derive the limiting
  value of Ohmic diffusion that is needed to inhibit the growth of the
  instability for different types of rotation laws. We also present
  an estimate of the level of turbulence needed to produce turbulent
  Ohmic diffusion strong enough to suppress MRI in the isotropic
  homogeneous case using the first-order smoothing approximation
  (hereafter FOSA). In Sect.~\ref{model}, we introduce the model
  that we use to study the dynamics of the MRI under the influence
  of external nonhelical forcing. First we apply this model to
  carry out simple one-dimensional calculations to reproduce the
  linear results of Sect.~\ref{LSA} and then to
  three-dimensional hydro- and magnetohydrodynamic models of
  isothermal, nonstratified, rotating and shearing cubic domains
  subject to rotation laws both unstable and stable with respect to
  the MRI. The latter have been used to calibrate the former to get a
  better understanding of turbulence resulting from the MRI with
  respect to the turbulence resulting from the interaction of external
  forcing, rotation and shear. The results of these calculations are
  presented in Sect.~\ref{res}. In Sect.~\ref{MW} we present a
  simple application of our results to galactic disks. In
  Sect.~\ref{conc} we present our conclusions.

\section{Linear stability analysis of non-ideal MRI}\label{LSA}

Jin (1996) presented linear stability analysis of non-ideal MRI for
disks with Keplerian rotation profiles. Here we generalise the theory
to include any rotation profile of the form $\Omega \propto r^{-q}$,
further assuming that the system is incompressible and pressureless.

\begin{figure}
\includegraphics[width=80mm,height=50mm]{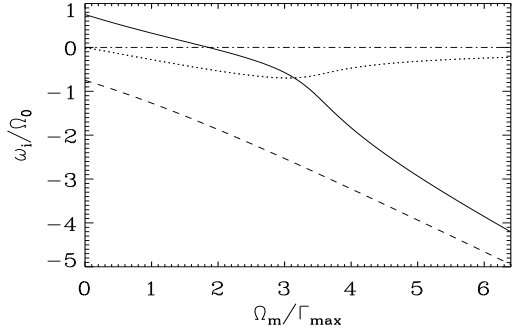} %idl/nmriroots.pro res/roots_q15.dat
\caption{Imaginary parts of the roots of the non-ideal dispersion relation for the Keplerian case $q=1.5$ as function of the Ohmic diffusion rate
  scaled with the maximum growth rate $\Gamma_{\rm max}$ of the MRI. Solid
  thick line represents the (fast) MRI mode, dashed line the decay
  mode, and the dotted line the slow mode
(given by two roots with equal imaginary parts). The dashed-dotted
  line indicates $\omega_i/\Omega_0$=0.}
\label{roots_q15}
\end{figure}

We choose a local Cartesian frame of reference, in which $x$, $y$ and $z$
denote the radial, toroidal and vertical directions, respectively. We
start by linearizing the pressureless ideal equation of motion
\begin{equation} \label{eq:equmotion}
\frac{\partial \bm{U}}{\partial t} + (\bm{U} \cdot \bm\nabla)\bm{U} +
2\,\bm{\Omega} \times \bm{U} - \frac{1}{\rho} \bm{J}\times\bm{B} = 0\;,
\end{equation}
and the non-ideal induction equation
\begin{equation} \label{eq:ind}
\frac{\partial \bm{B}}{\partial t} - \bm\nabla \times (\bm{U} \times \bm{B})
= \eta\bm\nabla^2 {\bm B}\;,
\end{equation}
around the location of the centre of the box, $r_0$, in a 
differentially rotating disk that is
subject to a weak uniform vertical magnetic field $B_0\hat{\bm e}_z$. Here $\bm{U}$ is the
velocity field, $\bm{B}$ is the magnetic field, 
$\bm{J}=\mu_0^{-1}\bm\nabla\times\bm{B}$ is the current density,
$\mu_0$ is the vacuum permeability, and
$\eta$ is the Ohmic diffusion coefficient. The angular velocity at $r_0$ is
$\Omega_0$, and it varies with the distance from the centre as $\Omega
\propto r^{-q}$. Such differential rotation 
translates into a radially linear shearing velocity ${\bm U}_0=-q
\Omega_0 x \hat{\bm e}_y$ after the linearization of
Eq.~(\ref{eq:equmotion}). Small fluctuations ${\bm
  u}=\left[u_x,u_y,u_z\right],\ {\bm b}=\left[b_x,b_y,b_z\right]$ are
now added to the equilibrium state ${\bm U}_0=\left[0,-q\Omega_0 x,0
  \right],\ {\bm B}_0=\left[0,0,B_0\right]$. We consider only
fluctuations in the vertical ($z$) direction, as fluctuations of this
type have been shown to have the highest growth rates (Balbus \&
Hawley\ 1992); in this case the equations for the $z$-components of
velocity and magnetic field decouple from the horizontal ones and
therefore they can be neglected. The set of linearised equations for the
horizontal fluctuating fields in component form read
\begin{eqnarray}\label{eq:lineqs}
\frac{\partial u_x}{\partial t}&=&2\,\Omega_0 u_y + \frac{B_0}{\mu_0 \rho}
\frac{\partial b_x}{\partial z}\;, \label{eq:1}\\
\frac{\partial u_y}{\partial t}&=&-\left(2 -q \right) \Omega_0 u_x +
\frac{B_0}{\mu_0 \rho} \frac{\partial b_y}{\partial z}\;,\label{eq:2} \\
\frac{\partial b_x}{\partial t}&=&B_0 \frac{\partial u_x}{\partial
z} + \eta \frac{\partial^2 b_x}{\partial z^2}\;,\label{eq:3} \\
\frac{\partial b_y}{\partial t}&=&-q \Omega_0 b_x + B_0 \frac{\partial
u_y}{\partial z} + \eta \frac{\partial^2 b_y}{\partial z^2}\;.\label{eq:4}
\end{eqnarray}
Let us now assume that the solutions to these equations are of the
form $f_j=\hat{f}_j e^{i \left( k z - \omega t \right)}$. Substitution
yields a set of algebraic equations
\begin{eqnarray}\label{eq:flineqs}
- i \omega \hat{u}_x &=&2\,\Omega_0 \hat{u}_y + i k \frac{B_0}{\mu_0 \rho}
\hat{b}_x\;, \\
- i \omega \hat{u}_y&=&-\left(2 -q \right) \Omega_0 \hat{u}_x + i k
\frac{B_0}{\mu_0 \rho} \hat{b}_y\;,\\
- i \omega \hat{b}_x&=& i k B_0 \hat{u}_x - k^2 \eta \hat{b}_x\;,\\
- i \omega \hat{b}_y&=&-q \Omega_0 \hat{b}_x + i k B_0 \hat{u}_y -k^2 \eta \hat{b}_y\;.
\end{eqnarray}
From these equations the following dispersion relation is obtained:
\begin{eqnarray}
\omega^{4} &+& 2 i \Omega_{m} \omega^{3} \nonumber \\ 
&-& \left( 2 \left( 2-q \right) \Omega_{0}^{2} + \Omega_{\rm m}^{2} + 2 k^{2} v_{\rm A}^{2} \right) \omega^{2} \nonumber \\
&-& i\Omega_{m}\left( 4 \left(2-q \right)\Omega_{0}^{2} + 2 k^{2} v_{\rm A}^{2} \right) \omega \nonumber \\
&+&\!k^{2} v_{\rm A}^{2}(k^{2} v_{\rm A}^{2}\!-\!2q \Omega_{0}^{2})
\!+\!2 \left( 2-q \right) \Omega_{0}^{2}\Omega_{\rm m}^{2}\!=\!0,
\end{eqnarray}
where $v^2_{\rm A}=\frac{B_0^2}{\mu_0 \rho}$, and $\Omega_{\rm m} = \eta k^2$. 
We have numerically solved the dispersion relation searching 
especially for the limiting Ohmic diffusion rate at which all the MRI
modes become damped. In practise we solve for all the four roots of
the dispersion relation and require that all the imaginary parts are
negative. The results of this analysis are plotted in 
Figs.~\ref{roots_q15} and \ref{LSA_damp}.

In Fig.~\ref{roots_q15} we show all imaginary parts of the four roots of the
dispersion relation for $q=1.5$, $k_{\rm max}/k=1$, where 
$k_{\rm max} =\frac{q}{2} \frac{v_{\rm A}}{\Omega_0} \sqrt{\frac{4}{q}-1}$,
as functions of the Ohmic diffusion rate $\Omega_{\rm m}$.
We have scaled $\Omega_{\rm m}$ with the
maximum growth rate, $\Gamma_{\rm max}$, of the ideal MRI. 
From these results it can be seen that in the ideal limit 
$\Omega_{\rm m}=0$ the growth rate of the MRI mode is positive
and equal to $\Gamma_{\rm max}=\frac{q}{2} \Omega_0$. As the diffusion is
increased, the growth rate decreases and reaches zero at
$\Omm\approx1.9 \Gamma_{\rm max}$. As the growth rates of the
other modes are negative for all values of $\Omega_{\rm m}$, then this is
the limiting diffusion rate above which the instability will be damped
for $k=k_{\rm max}$. Our results are in agreement with those of
Jin (1996); due to the incompressibility assumption the inertial wave
mode is absent in our analysis, but the limiting Ohmic diffusion rate
for MRI damping is identical for the Keplerian case in both studies.

In Fig.~\ref{LSA_damp} we have determined the limiting Ohmic
diffusion rate for each value of $q,\ 0 < q < 2$, where the system is ideally unstable, 
for all values of $k=[0,k_{\rm crit}]$, where
$k_{\rm crit}=\frac{\Omega_0}{v_{\rm A}}\sqrt{2 q}$ is the largest MRI-unstable
ideal wavenumber. Analytical `stability criterion' can also be derived
from the analysis: MRI will be damped if
\begin{equation}
\Omega_{\rm m} > \frac{k v_{\rm A}}{\Omega_0} \sqrt{\frac{2 q \Omega_0^2 - k^2 v_{\rm A}^2}{2 \left(2-q \right)}}.
\end{equation}
At the limit $q=2$, the Ohmic diffusion rate becomes infinite, meaning
that in this regime the flow cannot be stabilized by the Ohmic
diffusion. We find that for small $q$ the results for the maximally
excited and all unstable wavenumbers are identical, whereas for larger
$q$ they start to deviate more and more indicating that some
wavenumbers are harder to damp than the one with the largest growth
rate. From this plot we can also
see that for very small $q$ the limiting value approaches the maximum
growth rate, $\Gamma_{\rm max}$, of the ideal MRI. The limiting Ohmic
diffusion rate increases with $q$ above $\Gamma_{\rm max}$ so that for rotation profiles near the Rayleigh unstable regime ($q>2$)
the Ohmic diffusion rate needed to damp the MRI is roughly three times
$\Gamma_{\rm max}$.

\subsection{FOSA-prediction for isotropic homogeneous turbulence}

\begin{figure}
\includegraphics[width=80mm,height=50mm]{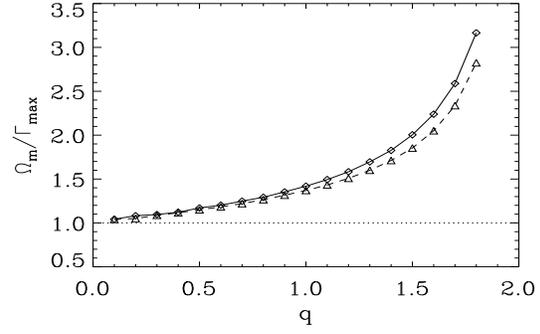} %idl/damp.pro res/damp.dat
\caption{Limiting Ohmic diffusion rate $\Omega_{\rm m}$ at which MRI
  modes corresponding to the range of wavenumbers $k=[0,k_{\rm crit}]$
  (solid line with diamonds) and for a single wavenumber $k=k_{\rm
    max}$ (dashed line with triangles) are damped as functions of
  $q$. The results are scaled with the maximum growth rate of the
  ideal MRI, $\Gamma_{\rm max}$. The dotted line shows the limit
  at which the Ohmic diffusion rate and maximum growth rate of MRI are
  equal for reference.}
\label{LSA_damp}
\end{figure}

Let us now use the information obtained from the linear stability
analysis to make a prediction for the nonlinear setup we use in the rest of the
paper. Firstly we need an estimate of the turbulent Ohmic diffusion
resulting from the isotropic external forcing at a wavenumber
$\kef$. In order to obtain this we can use the first-order smoothing 
approximation
(FOSA) result
\begin{equation}
\etat=\onethird \tau_c u_{\rm rms}^2 = \onethird u_{\rm rms} \kef^{-1}, \label{equ:etat}
\end{equation}
where we have assumed that the Strouhal number, ${\rm St}=\tau_c \urms
\kef$, is equal to unity.
This value has been confirmed in the kinematic regime numerically by
Sur et al.\ (2008) from non-shearing setups using the test field
method (Schrinner et al.\ 2005, 2007). In the presence of shear the
value of $\etat$ can increase slightly but is still of the same order
of magnitude (Brandenburg et al.\ 2008; Mitra et al.\ 2009).
Therefore we obtain
\begin{equation} \label{OmFOSA}
\Omega_{\rm m} = \etat \kef^2 = \onethird \urms \kef,
\end{equation}
for the turbulent Ohmic diffusion rate. From the linear stability
analysis we know that MRI will be damped if
\begin{equation} \label{dampcond}
\Omega_{\rm m} > A(q) \Gamma_{\rm max}= \frac{q A(q)}{2} \Omega_0,
\end{equation}
where $A(q)$ varies from 1 to roughly 3 as $q$ is increased from 0 to 2 (see
Fig.~\ref{LSA_damp}). By substituting the FOSA-expression, 
Eq.~(\ref{equ:etat}), we can
simplify this further, yielding the condition
\begin{equation}\label{LSApred}
\urms > \frac{3 q A(q) \Omega}{2 \kef},
\end{equation}
for the suppression of MRI. Note that for a given $\urms$ this condition 
is dependent on
the spatial scale of the turbulence so that the larger the scale, the
harder it is to damp MRI.

\section{Model}\label{model}

We use a model setup similar to that used in Liljestr\"om et
al.\ (2009), except that here we add an external forcing function in
the Navier--Stokes equation. The computational domain is a cube of size 
$(2\pi)^3$, filled with isothermal gas; vertical gravity and
stratification are neglected. The calculations are local and utilise the
shearing-sheet approximation.  The equations to be solved read
\begin{eqnarray}
\frac{\mathcal{D} \rho}{\mathcal{D} t} &=& - \bm{\nabla} \cdot (\rho \bm{u})\;, \\ \label{cont}
\frac{\mathcal{D} \bm{u}}{\mathcal{D} t}&=&-(\bm{u} \cdot \bm{\nabla})\bm{u} - q \Omega_0 u_x \hat{\bm{e}}_y - 2\,\Omega_0\hat{\bm{e}}_z \times \bm{u} \nonumber \\ && \hspace{0cm}-\frac{1}{\rho}\bm{\nabla} p + \frac{1}{\rho}\bm{J} \times \bm{B} 
+ \nu_{\rm h} \nabla^6 \bm{u} + f_{\rm force}, \\ \label{velocity}
\frac{\mathcal{D} \bm{A}}{\mathcal{D} t} &=& \bm{u} \times \bm{B} 
    + q \Omega_0 A_y \hat{\bm{e}}_x 
+ \eta_{\rm h} \nabla^6 \bm{A}\;, \label{eq:indlocal}
\end{eqnarray}
where $\mathcal{D} / \mathcal{D} t = \partial / \partial t + U_0
\partial / \partial y$ 
includes advection by the shear flow,
$\bm{u}$ is the departure from the mean flow $\bm{U}_0$, $\rho$ is the
density, $p$ is the pressure, $\bm{A}$ is the magnetic vector potential, $\bm{B} =
\bm{\nabla} \times {\bm A}$ is the magnetic field, and ${\bm J} =
\bm{\nabla} \times {\bm B}/\mu_0$ is the current density.
In order to get as close as possible to the ideal limit, we use
hyperviscous operators of the form $\nabla^6$ to replace the ordinary
viscosity $\nabla^2$-operators; $\nu_{\rm h}$ and $\eta_{\rm h}$ stand
for the hyperviscous kinematic viscosity and magnetic diffusivity,
respectively. With this procedure we aim at maximizing the Reynolds
number in the quiescent regions of the flow while diffusing and
damping fluctuations near the grid scale. Compared to direct
simulations (e.g.\ Haugen \& Brandenburg 2006) with uniform
viscosities, smaller grid size can be used to resolve the flow.

The forcing function $\bm{f}_{\rm force}$ is given by
\begin{eqnarray} \label{ff}
\bm{f}(\bm{x},t) = {\rm Re} \{N \bm{f}_{\bm{k}(t)} \exp [i \bm{k}(t)
  \cdot \bm{x} - i \phi(t) ] \}\;,
\end{eqnarray}
where $\bm{x}$ is the position vector, $N = f_0 c_{\rm s} (k c_{\rm
  s}/\delta t)^{1/2}$ is a normalization factor, $f_0$ is the forcing
amplitude, $k = |\bm{k}|$, $\delta t$ is the length of the time step,
and $-\pi < \phi(t) < \pi$ a random delta-correlated phase. The vector
$\bm{f}_{\bm{k}}$ is given by
\begin{eqnarray}
\bm{f}_{\ve{k}} = \frac{\bm{k} \times \hat{\bm{e}}}{\sqrt{\bm{k}^2 - (\bm{k}
    \cdot \hat{\bm{e}})^2}}\;,
\end{eqnarray}
where $\hat{\bm{e}}$ is an arbitrary unit vector. Thus, $\bm{f}_{\bm{k}}$
describes nonhelical transversal waves with $|\bm{f}_{\ve{k}}|^2 = 1$,
where ${\bm k}$ is chosen randomly from a predefined range in the vicinity
of the average nondimensional wavenumber $\kef/k_1$ at each time
step. Here $k_1$ is the wavenumber corresponding to the domain size,
and $\kef$ is the wavenumber of the energy-carrying scale.

Periodic boundary conditions are applied in all three directions; in
the radial direction we account for the shear flow $\bm{U}_0$ by making use
of the shearing-sheet approximation (e.g. Wisdom \& Tremaine 1988):
\begin{equation}
f(\onehalf L_x,y,z)=f(-\onehalf L_x,y+q \Omega_0 L_x t,z),
\end{equation}
where $f$ stands for any of the seven independent variables, $L_x$ for
the radial extent of the computational domain, and $t$ is the time.

The domain is initially threaded by a weak magnetic field,
\begin{equation}
{\bm A} = A_0 \hat{\bm{e}}_y \cos k_{\rm A}x \cos k_{\rm A}y \cos k_{\rm A}z\;.
\end{equation}
Thus, the magnetic field contains periodic $x$- and $z$-components
with amplitude $A_0$. The values of $k_{\rm A}$, $\Omega_0$ and $A_0$
are selected so that both the wavenumber with the largest growth rate,
$k_{\rm max}$, and the largest unstable wavenumber, $k_{\rm crit}$,
are well resolved by the grid; in practise this means that in the three-dimensional calculations we
adopt $k_{\rm A}/k_1=1$, $\Omega_0 = 0.2 c_{\rm s}k_1$ and $A_0 = 0.2
\sqrt{\mu_0\rho_0}\,c_{\rm s}k_1^{-1}$, resulting in $k_{\rm max} =
O(k_1)$. For the initial setup, the other condition for the onset of
MRI, namely $\beta \gg 1$, where $\beta=2\mu_0 p/B_0^2$ is
the ratio of the thermal to magnetic pressure, is also satisfied as
$\beta$ is at least 50 at the maximum values of the magnetic
field. 

For all the simulations we use the {\sc
  Pencil Code}\footnote{\textsf{http://www.nordita.org/software/pencil-code/}},
which is a high-order (sixth order in space, third order in time),
finite-difference code for solving the MHD equations (e.g.\ Brandenburg \&
Dobler 2002).

The local calculations have been carried out at two different
resolutions, namely $128$ (in 1D) and $64^3$ (in 3D). In the
three-dimensional calculations the hyperdiffusion coefficients
used are $\nu_{\rm h} = \eta_{\rm h} = 2 \times 10^{-7} c_{\rm s} k_1^{-1}$.

\section{Results}\label{res}

\subsection{Linear regime}\label{1Druns}

\begin{figure}
\includegraphics[width=80mm,height=50mm]{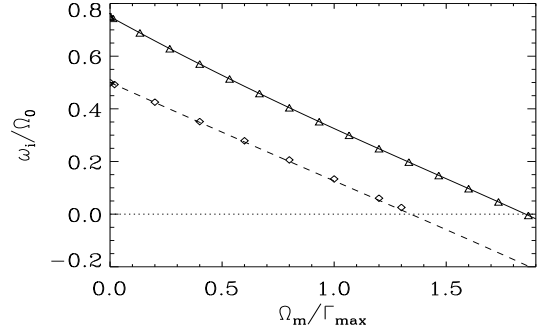} %idl/lin.pro res/q15_damp.dat res/q1_damp.dat res/roots_q15.dat res/roots_q1.dat
\caption{The growth rate $\omega$ of the MRI-mode with $k_{\rm max}=k_1$,
  $\Omega_0/(c_{\rm s} k_1)=1$, and $v_{\rm A}/c_{\rm s}=1$
  as a function of $\Omega_{\rm m}$. Solid line: linear stability
  result for the Keplerian rotation law with $q=3/2$. Dashed line:
  the same for flat galactic rotation curves with $q=1$. Triangles and
  diamonds show the growth rates measured from the one-dimensional
  numerical model.}
\label{lin_gammas}
\end{figure}

To validate the numerical method used in the three-dimensional
nonlinear calculations, we first try to reproduce the linear results of
Sect.~\ref{LSA} with a one-dimensional setup 
using ordinary magnetic diffusion with 
$\nabla^2$-operator and a constant coefficient $\eta$ instead of the
hyperdiffusion scheme.  For these runs we choose $k_{\rm max}=k_1$,
$\Omega_0/(c_{\rm s} k_1)=1$, $v_{\rm A}/c_{\rm s}=1$, vary $q$
and $\eta$, and impose a uniform magnetic field of strength
$B_0/(\sqrt{\rho_0 \mu_0} c_{\rm s})=\frac{q}{2} \sqrt{\frac{4}{q}-1}$
in the vertical ($z$) direction at each time step. This procedure
forces the system to develop MRI at the wavenumber $k=k_{\rm max}$
with the growth rate $\Gamma_{\rm max}$ at the ideal limit
($\Omega_{\rm m}=0$).  In Fig.~\ref{lin_gammas} we plot the growth
rate of the MRI mode as function of $\Omega_{\rm m}/\Gamma_{\rm max}$
for two values of $q$.  As can be seen from this figure, the results
of both methods are in perfect agreement.

\begin{figure}
\includegraphics[width=80mm,height=50mm]{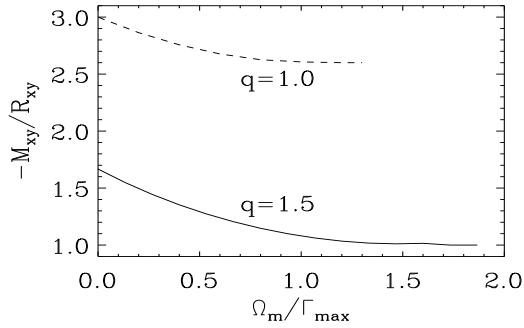} %idl/lin.pro res/q15_damp.dat res/q1_damp.dat res/roots_q15.dat res/roots_q1.dat
\caption{The Maxwell to Reynolds stress ratio $-M_{xy}/R_{xy}$ as a
  function of $\Omega_{\rm m}$ obtained from
  the one-dimensional calculations with Keplerian rotation
  law ($q=3/2$, solid line) and with flat rotation curve ($q=1$, dashed line).}
\label{1D-ratios}
\end{figure}

We also compute the time averages of the Maxwell and Reynolds
stresses, $M_{xy}= \mu_0^{-1}\left< b_x b_y \right>$ and
$R_{xy}=\left< u_x u_y\right>$, from these calculations.  In
the\ limit $\Omm=0$ the linear stress ratio reads
\begin{equation}\label{eq:lin-ratio}
-\frac{M_{xy}}{R_{xy}} = \frac{4-q}{q},
\end{equation}
as derived by Pessah et al.\ (2006). As can be seen from
Fig.~\ref{1D-ratios}, where we plot the stress ratio as a function of
$\Omm$ for the Keplerian ($q=3/2$) and flat ($q=1$) rotation laws from the
one-dimensional calculations, our results agree with
Eq.~(\ref{eq:lin-ratio}) for $\Omm=0$. As we increase $\eta$ and
thereby $\Omm$, the stress ratios decrease, reaching minima somewhat
before the limiting Ohmic diffusion rates above which the MRI is
damped and the growth rates are negative.  Although the stress ratios
in the fully nonlinear state of the MRI differ from the linear ones
(see e.g. Liljestr\"om et al.\ 2009), these results suggest that the
sress ratio should show a decreasing trend as a function of the
turbulent Ohmic diffusion rate also in the nonlinear regime.

\subsection{Three-dimensional calculations with external forcing}

Let us now consider the implications of the above analysis to the
earlier numerical work done by Workman \& Armitage (2008). They used a
setup very similar to the one presented in Sect.~\ref{model},
varying the amplitude $f_0$ and the energy carrying wavenumber $\kef$
of the forcing function given by Eq.~(\ref{ff}). Two different forcing
scales were investigated, namely a large-scale forcing with
$\kef/k_1=1.5$ and a small-scale forcing with $\kef/k_1=40$, the other
relevant parameters chosen to be $\Omega_0/(c_{\rm s} k_1)=0.2$ and 
$q=1.5$. For these
parameters our analysis predicts roughly a factor of 30 difference in
the limiting rms-values, the exact numbers reading $\urms/c_s =
0.6$ for the large-scale forcing, and $\urms/c_s = 0.0225$ for
the small-scale one. Workman \& Armitage (2008) report a running time
average of the kinetic energy density $E_{\rm kin}=0.0012$ for the
pure MRI case, and roughly 15 times more kinetic energy for the
largest forcing amplitude investigated.  Assuming that the average
density is roughly one, the rms-velocities would read
$\urms/c_s\approx0.049$ and 0.190, for the pure and maximally forced
runs, respectively. Based on this estimate, all of the cases with the
large-scale forcing investigated by Workman \& Armitage (2008)
produced turbulence whose level was below the suppression limit; we
propose that this is the reason why no clear MRI damping was seen with
large-scale forcing. On the other hand, considering the small-scale forcing, the
limiting rms-velocity is smaller than the one of pure MRI, so MRI
should be suppressed at any forcing amplitude.

We perform four sets of three-dimensional nonlinear calculations
differing by the effective forcing wavenumber in the range
$\kef/k_1=[1.5,3,5,10]$. The parameters common to all runs are 
$\Omega_0/(c_{\rm s}k_1)=0.2$, $q=1.5$, and $A(q) \approx 2$. The
predicted limiting rms-velocities for the chosen wavenumbers are
$u_{\rm lim}/c_{\rm s}=[0.6,0.3,0.18,0.09]$. In the
following we calibrate the forcing amplitudes of the calculations to
produce turbulent rms-velocities below and above these limiting values.

\subsubsection{Hydrodynamic calculations}\label{hydro}

\begin{table}
 \centering%%%
\caption{Hydrodynamic runs with $64^3$ mesh points, with averages calculated
  over 150 rotations. 
  Rotation profile with $q=3/2$ i.e. angular velocity decreasing 
  outwards was used in these calculations. The number in the brackets
  in the third column indicates the power on ten.}
\label{hydro_runs}
\begin{tabular}{llcc}\hline
&$f_0$ &$\brac{u_xu_y}/c_{\rm s}^2$ &$u_{\rm rms}/c_{\rm s}$\\
\hline
$\kef/k_1=1.5$ &0.01  &-1.91 (-6) &0.07\\ %k1_hydro.dat
           &0.025 & 3.35 (-4) &0.14\\
           &0.05  & 2.25 (-3) &0.25\\
           &0.10  & 7.90 (-3) &0.40\\
           &0.15  & 1.36 (-2) &0.52\\
           &0.20  & 2.07 (-2) &0.63\\
           &0.25  & 2.70 (-2) &0.72\\
           &0.27  & 2.94 (-2) &0.76\\
\hline
$\kef/k_1=3$ &0.01   &3.15 (-5) &0.07 \\ %k3_hydro.dat
         &0.025  &8.62 (-4) &0.16 \\
         &0.05   &3.69 (-3) &0.27 \\
         &0.10   &1.04 (-2) &0.42 \\
         &0.15   &1.70 (-2) &0.54 \\
         &0.20   &2.36 (-2) &0.64 \\
         &0.25   &2.81 (-2) &0.72 \\
\hline
$\kef/k_1=5$ &0.01  &6.00 (-5) &0.07\\ %k5_hydro.dat
         &0.025 &9.98 (-4) &0.16\\
         &0.05  &3.64 (-3) &0.26\\
         &0.10  &9.38 (-3) &0.41\\
         &0.15  &1.41 (-2) &0.52\\
         &0.20  &1.84 (-2) &0.61\\
\hline
$\kef/k_1=10$ &0.01  &6.28 (-5) &0.06\\ %k10_hydro.dat
          &0.025 &6.24 (-4) &0.13\\
          &0.05  &2.22 (-3) &0.21\\
          &0.10  &5.38 (-3) &0.34\\
          &0.15  &7.77 (-3) &0.45\\
\hline
\end{tabular}
\end{table}

In a related study (Snellman et al.\ 2009), the Reynolds stresses in 
forced turbulence with rotation and shear were investigated; a non-zero
$R_{xy}$-component naturally arises in such a setup. This is in
disagreement with the findings of Workman \& Armitage\ (2008), who
report very small $R_{xy}$, interpreted to be consistent with zero in
their hydrodynamic reference runs. This result was taken to show that
the hydrodynamic forcing alone cannot produce any stress and that all
the stresses in
the magnetohydrodynamic runs were due to the MRI and its interaction with
the background turbulence.
This discrepancy
between the two very similar studies prompted us to perform a series
of hydrodynamic runs with varying forcing wavenumber and amplitude and
monitor the time evolution of the Reynolds stress component
$R_{xy}$. 
Four sets of runs with $\kef/k_1=[1.5,3,5,10]$ were
performed. 
Our results are summarised in Table~\ref{hydro_runs}. As is
evident from this table, we observe a non-zero $R_{xy}$ in all our
calculations, the magnitude increasing with forcing amplitude
within each set. No great qualitative differences are found between the sets with
varying forcing wavenumber.
These results clearly indicate that in the runs with both external
forcing and MRI, non-zero Reynolds and Maxwell stresses can be
expected to be generated both from the forcing and the MRI, and the
two contributions should be carefully separated.

In Snellman et al.\ (2009) it is shown that for a given $q$, the
Reynolds stresses vary as a functions of $\Omega$ and $S$, see
especially their Fig.\ 11. In the hydrodynamic run presented by
Workman \& Armitage (2008) the Coriolis number, ${\rm
  Co}=2\Omega/(\urms \kef)$, measuring the effect of rotation, is
roughly 5.3. The results of Snellman et al.\ (2009) indicate that in
this regime the stress $R_{xy}$ is small and negative although the
runs are probably not directly comparable due to the different forcing
scales. The conclusion of Workman \& Armitage (2008) that hydrodynamic
stresses are small is based on this one point, which just by chance
lies in an unfortunate part of the parameter space.

\begin{figure}
\includegraphics[width=80mm,height=50mm]{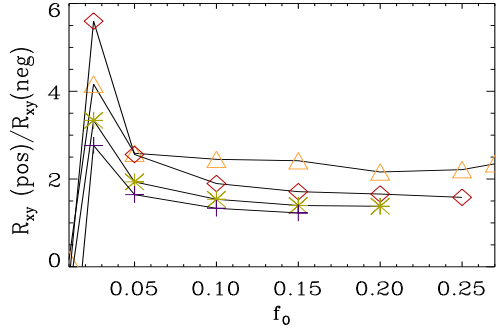} %idl/pnneg.pro
\caption{Ratio of the Reynolds stresses $R_{xy}$ with positive shear
  parameter to the ones with negative shear parameters from the
  hydrodynamic three-dimensional runs as function of forcing amplitude
  $f_0$. Orange color and triangles: $\kef/k_1=1.5$, red color and
  diamonds: $\kef/k_1=3$, green color and stars: $\kef/k_1=5$,
  and violet color and crosses: $\kef/k_1=10$.}
\label{hydro_asymm}
\end{figure}

In the study of Snellman et al.\ (2009) both positive ($\Omega$
decreasing outwards) and negative ($\Omega$ increasing outwards) shear
parameters $q$ were investigated systematically. It was found that the
stresses were not symmetric with respect to $q$ in the sense that when
changing $q \rightarrow -q$ does not mean that the magnitude of the
stress remains the same.  The same behaviour is expected to occur in
our setup; to test this, we produced sets of calculations identical to
the ones presented in the previous paragraph but with negative shear
parameter of the same magnitude, namely $q=-3/2$. The results are
summarised in Table~\ref{hydro_runs_neg}.  Comparing these numbers to
the ones listed in Table~\ref{hydro_runs}, it can be seen that a
significant asymmetry indeed exists between the positive and negative
shear parameter runs, and that this asymmetry is strongly dependent on
the wavenumber. As can be seen from Fig.~\ref{hydro_asymm}, for the
largest wavenumber ($\kef/k_1=10$) investigated, the asymmetry is
quite weak (the runs with positive shear parameter produce Reynolds
stresses only slightly larger than the negative shear parameter runs)
whereas for the smallest wavenumber, $\kef/k_1=1.5$, the ratio of
positive to negative shear parameter-generated Reynols stress is more
than two.

\begin{table}
 \centering%%%
\caption{The same as Table~\ref{hydro_runs}, but rotation profile with $q=-3/2$, i.e. angular velocity increasing outwards, was used.}
\label{hydro_runs_neg}
\begin{tabular}{llcc}\hline
&$f_0$ &$\brac{u_xu_y}/c_{\rm s}^2$ &$u_{\rm rms}/c_{\rm s}$\\
\hline
$\kef/k_1=1.5$ &0.01  &-1.15 (-5) &0.09\\ %k1_hydro_neg.dat
           &0.025 & 8.06 (-5) &0.14\\
           &0.05  & 8.69 (-4) &0.23\\
           &0.10  & 3.22 (-3) &0.38\\
           &0.15  & 5.62 (-3) &0.49\\
           &0.20  & 9.59 (-3) &0.59\\
           &0.25  & 1.22 (-2) &0.68\\
           &0.27  & 1.25 (-2) &0.72\\
\hline
$\kef/k_1=3$ &0.01   &-9.02 (-5) &0.07\\ %k3_hydro_neg.dat
         &0.025  & 1.54 (-4) &0.15\\
         &0.05   & 1.44 (-3) &0.26\\
         &0.10   & 5.48 (-3) &0.41\\
         &0.15   & 9.91 (-3) &0.53\\
         &0.20   & 1.42 (-2) &0.62\\
         &0.25   & 1.78 (-2) &0.71\\
\hline
$\kef/k_1=5$ &0.01   &-9.15 (-5) &0.07\\ %k5_hydro_neg.dat
         &0.025  & 2.99 (-4) &0.16\\
         &0.05   & 1.88 (-3) &0.26\\
         &0.10   & 6.10 (-3) &0.40\\
         &0.15   & 1.01 (-2) &0.51\\
         &0.20    & 1.33 (-2) &0.60\\
\hline
$\kef/k_1=10$ &0.01  &-1.96 (-5)    &0.06\\ %k10_hydro_neg.dat
          &0.025 & 2.26 (-4)    &0.13\\
          &0.05  & 1.35 (-3)    &0.21\\
          &0.10  & 4.04 (-3)    &0.34\\
          &0.15  & 6.33 (-3)    &0.45\\
\hline
\end{tabular}
\end{table}

\begin{figure}
\includegraphics[width=80mm,height=50mm]{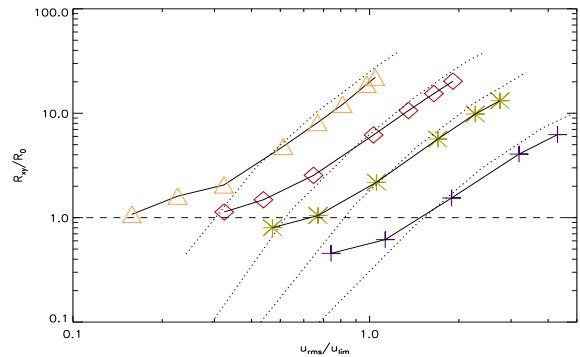}
\caption{Reynolds stresses $R_{xy}$ generated in the MRI-unstable runs
  with varying forcing wavenumber and forcing amplitude, normalised to
  $R_0$, the Reynolds stress generated in the pure MRI run wihtout
  forcing. Orange color and triangles: $\kef/k_1=1.5$, red color and
  diamonds: $\kef/k_1=3$, green color and stars: $\kef/k_1=5$, and
  violet color and crosses: $\kef/k_1=10$. The dotted lines plotted
  over each curve indicate the corresponding result from purely hydrodynamic
  calculations with identical normalisation.}
\label{rey_pure}
\end{figure}

\begin{figure}
\includegraphics[width=80mm,height=50mm]{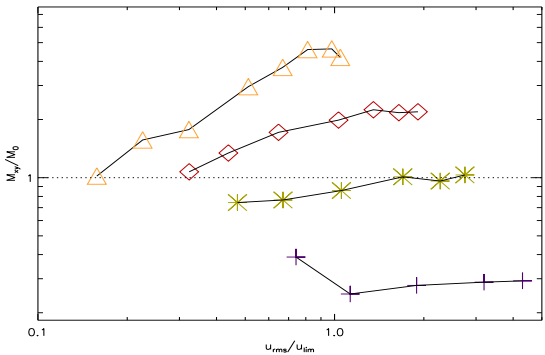}
\caption{Maxwell stresses $M_{xy}$ generated in the MRI-unstable runs
  with varying forcing wavenumber and forcing amplitude, normalised to
  $M_0$, the Maxwell stress generated in the pure MRI run wihtout
  forcing. Colors and symbols as in Fig.~\ref{rey_pure}.}
\label{max_pure}
\end{figure}

\begin{table*}
 \centering%%%
\caption{MRI-unstable runs with nonhelical forcing with resolution
  $64^3$, averages calculated over 250 rotations. 
  The mean magnetic field 
  $\brac{B}=(\brac{\mean{B}_x^2}^{1/2}+\brac{\mean{B}_y^2})^{1/2}$, 
  where the angular brackets denote time averages and overbar 
  horizontal averages.}
\label{nonhel64}
\begin{tabular}{llcccccccccc}\hline
$\kef/k_1$ &$f_0$ &$\brac{u_xu_y}$ &$\mu_0^{-1}\brac{b_xb_y}$ &$-M_{xy}/R_{xy}$ &$u_{\rm rms}/c_{\rm s}$ &$B_{\rm rms}/(\sqrt{\mu_0 \rho_0} c_{\rm s})$ &$\brac{B}/B_{\rm rms}$\\ \hline
Pure MRI &-     &7.46 (-4)  &-3.27 (-3) &4.38 &0.08 &0.12 &0.08\\ %1.5nofor
\hline
1.5 &0.01  &8.03 (-4) &-3.33 (-3) &4.15 &0.09 &0.12 &0.36\\
    &0.025 &1.20 (-3) &-5.13 (-3) &4.28 &0.14 &0.15 &0.29\\
    &0.05  &1.54 (-3) &-5.81 (-3) &3.78 &0.19 &0.17 &0.35\\
    &0.10  &3.60 (-3) &-9.68 (-3) &2.69 &0.31 &0.22 &0.23\\
    &0.15  &6.08 (-3) &-1.21 (-2) &2.00 &0.40 &0.26 &0.24\\
    &0.20  &9.08 (-3) &-1.51 (-2) &1.66 &0.49 &0.30 &0.21\\
    &0.25  &1.41 (-2) &-1.52 (-2) &1.07 &0.59 &0.32 &0.21\\
    &0.27  &1.65 (-2) &-1.37 (-2) &0.83 &0.63 &0.31 &0.21\\
\hline
3 &0.01  &8.46 (-4)  &-3.51 (-3) &4.15 &0.10 &0.12 &0.31\\ %1.5nohelfor0.01
  &0.025 &1.11 (-3)  &-4.39 (-3) &3.96 &0.13 &0.14 &0.30\\ %1.5nohelfor0.025
  &0.05  &1.90 (-3)  &-5.62 (-3) &2.96 &0.19 &0.16 &0.28\\ %1.5nohelfor0.05
  &0.10  &4.62 (-3)  &-6.50 (-3) &1.41 &0.31 &0.20 &0.25\\ %1.5nohelfor0.1
  &0.15  &7.92 (-3)  &-7.35 (-3) &0.93 &0.41 &0.24 &0.17\\ %1.5nohelfor0.15
  &0.20  &1.15 (-2)  &-7.10 (-3) &0.62 &0.49 &0.27 &0.18\\ %1.5nohelfor0.2
  &0.25  &1.51 (-2)  &-7.17 (-3) &0.47 &0.57 &0.29 &0.16\\ %1.5nohelfor0.25
\hline
5 &0.01   &5.99 (-4) &-2.43 (-3) &4.06 &0.08 &0.10 &0.31 \\
  &0.025  &7.82 (-4) &-2.51 (-3) &3.21 &0.12 &0.11 &0.32\\
  &0.05   &1.63 (-3) &-2.81 (-3) &1.72 &0.19 &0.13 &0.24\\
  &0.10   &4.23 (-3) &-3.31 (-3) &0.78 &0.31 &0.17 &0.16\\
  &0.15   &7.34 (-3) &-3.14 (-3  &0.43 &0.41 &0.20 &0.16\\
  &0.20   &9.83 (-3) &-3.38 (-3) &0.34 &0.50 &0.23 &0.13\\
\hline
10 &0.01  &3.40 (-4) &-1.27 (-3) &3.74 &0.07 &0.08 &0.36\\
   &0.025 &4.60 (-4) &-8.18 (-4) &1.78 &0.10 &0.07 &0.41\\
   &0.05  &1.15 (-3) &-9.06 (-2) &0.79 &0.17 &0.09 &0.36\\
   &0.10  &3.03 (-3) &-9.42 (-2) &0.31 &0.29 &0.12 &0.20\\
   &0.15  &4.66 (-3) &-9.57 (-2) &0.21 &0.39 &0.14 &0.15\\
\hline
\end{tabular}
\end{table*}

\subsubsection{MRI-unstable runs}

Next we turn to the magnetohydrodynamic regime and investigate a
MRI-unstable setup with $k_{\rm A}/k_1=1$, $\Omega_0 = 0.2 c_{\rm
  s}k_1$, $A_0 = 0.2 \sqrt{\mu_0\rho_0}\,c_{\rm s}k_1^{-1}$, and
$q=3/2$. For reference, we make a pure MRI-run without forcing; the
results from this run are summarised in the first row of
Table~\ref{nonhel64}. The typical signatures of MRI (see
e.g.\ Liljestr\"om et al.\ 2009), namely a stress ratio of roughly
four and magnetic energy exceeding the kinetic energy, are
visible. The system does not exhibit large-scale dynamo action, so no
significant mean magnetic fields are present.

\begin{figure}
\includegraphics[width=80mm,height=50mm]{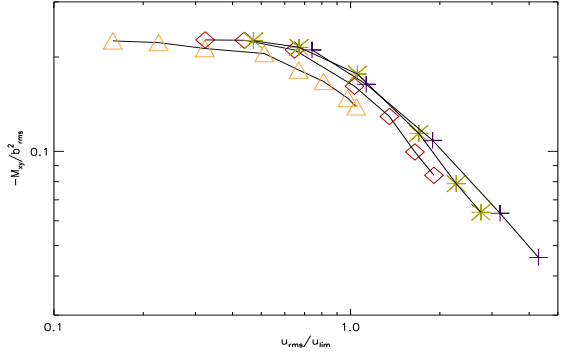}
\caption{Maxwell stresses $-M_{xy}$ generated in the MRI-unstable runs
  with varying forcing wavenumber and forcing amplitude, normalised to
  $b^2_{\rm rms}$ of each run. Colors and symbols as in
  Fig.~\ref{rey_pure}.}
\label{max_rms}
\end{figure}

\begin{figure}
\includegraphics[width=80mm,height=50mm]{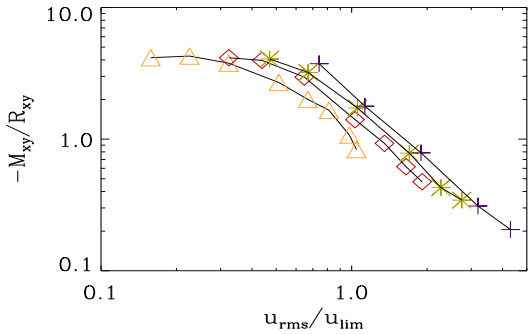}
\caption{Maxwell to Reynolds stress ratio $-M_{xy}/R_{xy}$ as function
  of $u_{\rm rms}$ from MRI-unstable runs with varying forcing
  wavenumber and amplitude. Colors and symbols as in Fig.~\ref{rey_pure}.}
\label{ratio}
\end{figure}

As in the hydrodynamic case, we proceed by making four sets of runs
with varying forcing wavenumber
and amplitude; the parameters for these runs are $\kef/k_1=[1.5,3,5,10]$
plotted in Figs.~\ref{rey_pure}-\ref{max_neg} with $[$violet crosses, green stars, red diamonds, orange
  triangles$]$, respectively.
Forcing amplitudes $f_0$ are selected for each set separately so that
turbulence levels below and above the limiting rms-velocities for MRI
damping $u_{\rm lim}/c_{\rm s}=[0.6,0.3,0.18,0.09]$ are obtained at
each $\kef$; the values used and the results summarised can be found
from Table~\ref{nonhel64}.

In Sect.~\ref{hydro} it was shown that purely hydrodynamic calculations
clearly produce non-zero Reynolds stresses due to
shear, rotation and the adopted forcing function. Therefore, in the
magnetohydrodynamic regime non-zero Maxwell stresses can also be
expected to be generated by the interaction of a small-scale dynamo
and shear
without the MRI. Therefore, the stresses in a
MRI-unstable setup can be written as
\begin{eqnarray}
R_{xy}&=&R_{xy}^{(\rm f)} + R_{xy}^{(\rm MRI)}, \\ 
M_{xy}&=&M_{xy}^{(\rm f)} + M_{xy}^{(\rm MRI)}.
\end{eqnarray}
Here we note that in the study of Workman \& Armitage\ (2008) $R_{xy}^{(\rm
  f)} = M_{xy}^{(\rm f)}=0$ was apparently assumed in the analysis of the results. To
find a definite answer to whether MRI is damped or not in the
investigated setup, therefore, a techinique of separating the stresses
should be developed. Before attempting that, let us first analyse the
behavior of the total stresses $R_{xy}$ and $M_{xy}$ measured in our
calculations.

The Reynolds stresses of the MRI-unstable calculations can well be
compared to the forced hydrodynamic ones of Sect.~\ref{hydro}; this is
done in Fig.~\ref{rey_pure}, where we plot both the
magnetohydrodynamic stresses (solid lines with symbols) and the
hydrodynamic stresses (dotted lines) normalised to the pure MRI
stress.  From this figure it can be seen that for the sets
$\kef/k_1=5$ and $10$, the weakest forcing amplitudes produce Reynolds
stresses that are smaller in magnitude than the pure MRI-generated
stress; this indicates that the MRI becomes reduced when external
forcing is added.  It is also evident that for the weakest forcings,
extra stresses in addition to the hydrodynamic runs are generated by
the action of MRI (the solid lines run above the dotted lines). As
the strength of the turbulence grows with increasing forcing, the
magnetohydrodynamic stress grows much more slowly than the
hydrodynamic one, so that the two curves start approaching each
other. This could be interpreted as follows: the action of the forcing
on the MRI-dynamics is to decrease its contribution $R_{xy}^{(\rm
  MRI)}$ to the total Reynolds stress, while $R_{xy}^{(\rm f)}$
continues its growth. At the strongest forcings and the highest levels
of turbulence, the hydrodynamic stress is always larger than the
magnetohydrodynamic one, but the shape of the curves is almost
identical; this could be interpreted as the MRI-contribution being
nearly zero while $R_{xy}^{(\rm f)}$ continues its growth. These
results are consistent with the scenario that the MRI becomes damped
as forcing is increased, but as the curves significantly deviate from
each other at high turbulence levels, this method cannot be used to
definitely determine at which point the damping occurs.  The stresses
from hydrodynamic runs are consistently higher than those from the
runs with MRI and forcing. A possible explanation is that the
backreaction due to the Lorentz force quenches the Reynolds stress in
the latter runs.  A weak $k$-dependence can be observed: the larger
the forcing wavenumber is, the smaller is the Reynolds stress which
can be interpreted as an effect of the decreasing influence of shear
and rotation on the system.

In contrast to the pure MRI case, in all the forced runs large-scale
dynamo action develops; such a shear dynamo has been previously
reported in rotating and shearing flows e.g.\ by Yousef et
al.\ (2008). Due to this a mean magnetic field is generated, showing
quite irregular behaviour in time disappearing and reappearing without
any cyclicity. The strength of the mean magnetic field is maximally
around 40 percent of the total magnetic field strength in the set
$\kef/k_1=10$, and minimally of the order of 10 percent with the
highest forcings. Such a mean field gives a significant additional
contribution to the Reynolds and Maxwell stresses. We remove this
contribution using
\begin{equation}
\brac{b_xb_y}=\brac{\overline{M}_{xy}^{(\rm total)}} - \brac{\overline{B}_x}\brac{\overline{B}_y},\label{removemf}
\end{equation}
where the overbars denote horizontal averaging.

For the Maxwell stress no such reference point exists as for the
Reynolds stress. In Fig.~\ref{max_pure} we simply plot the stress
normalised to the pure MRI case.  The behaviour of the stress is quite
similar for the three smallest wavenumber sets: first the stress is
growing as the forcing amplitude and turbulence level is increasing.
At around $u_{\rm lim}$, the growth saturates. For the highest
wavenumber set, the stress is at first strongly decreasing, but
saturation is again reached around $u_{\rm lim}$.  For the largest
wavenumber set, the total Maxwell stress is always below the pure MRI
case; also for the set $\kef/k_1=5$ the stress is weaker than the pure
MRI stress for the weakest forcings. This is a clear indication of the
MRI-contribution to the Maxwell stress being reduced when forcing is
included. If it is further assumed that the forcing-induced Maxwell
stress grows monotonically as function of forcing, then the combined
effect with a monotonically decreasing MRI-contribution at higher
forcing amplitudes could result in the saturation of the stress; this,
however, is only speculative, as no reference point is yet available.
A stronger $k$-dependence compared to the Reynolds stress can be seen
in the Maxwell stress: the larger the wavenumber, the smaller is the
stress. There is roughly an order of magnitude difference between the
Maxwell stress of $\kef/k_1=1.5$ and $10$.  This could be due to the
decreasing value of the effective magnetic Reynolds number, ${\rm
  Rm}=\urms/(\eta \kef)$.

The energy contained in the shear flow is pumped into the magnetic
energy reservoir by the Maxwell stresses; at the same time the
magnetic field is being sustained by the energy from the turbulent
forcing via small-scale dynamo action.  As long as the MRI is capable
of pumping the energy from the shear flow, the Maxwell stress
normalised with the square of the rms magnetic field strength should
remain more or less constant.  If the MRI-activity was reduced due to
the external forcing, indications of which were presented in the
previous paragraph, $b_{\rm rms}$ would be expected to continue growing
while the Maxwell stress gets suppressed; in Fig.~\ref{max_rms} we
plot the Maxwell stress normalised with $b^2_{\rm rms}$ as function of
rms-velocity. For the sets $\kef/k_1=[1,3,5]$ the normalised Maxwell
stress is indeed constant for turbulence levels weaker than the
LSA-predicted limiting value, but starts decreasing strongly for
stronger turbulence. For the set $\kef/k_1=10$ the Maxwell stress
shows a monotonically decreasing trend, as even the weakest forcing
cases are very close to $u_{\rm lim}$.

Keeping in mind the behavior of the individual stresses (Reynolds
stress monotonically growing while Maxwell stress saturates as the
forcing level is increased), one would expect the ratio of these
stresses, $-M/R$, to start strongly decreasing as the level of
turbulence exceeds $u_{\rm lim}$.  As can be seen from
Fig.~\ref{ratio}, where we plot the stress ratio observed in each set
of our calculations, the decrease of the stress ratio starts somewhat
before $u_{\rm lim}$, and occurs at the same exponential rate for all
the wavenumbers investigated. For the set $\kef/k_1=10$, for which the
widest range of $u_{\rm rms}/u_{\rm lim}$ is available, an order of
magnitude decrease is observed when the weakest and strongest forcings
are compared. Comparing these findings to the linear results with
ordinary Ohmic diffusion $\eta$ (Sect.~\ref{1Druns}), indicating that
the stress ratio decreases as $\eta$ increases, this behavior is
indicative of MRI suppression.

To summarise, comparing the stresses developing in the forced,
rotating and shearing runs to the pure MRI case does not directly give
information on the suppression or enhancement of the MRI, as also the
external forcing gives rise to non-zero stresses. A more appropriate
way of interpreting the stresses is either to normalise them with the
rms-values realised in the calculations, or to study the behavior of
the stress ratio. With the help of all these diagnostics, it seems
evident that the angular momentum transport due to MRI is reduced by
the turbulence from external forcing, and that the functional
dependence on the wavenumber seems to follow the LSA-FOSA
prediction. To definitely conclude that the MRI is totally damped, a
method of separating the forcng versus MRI-generated stresses should
be found. This is tried in the next subsection.

\begin{table*}
 \centering%%%
 \caption{Nonhelically forced $64^3$ runs without MRI (shear profiles
   with angular velocity increasing outwards. $B_{\rm rms}$ stands for
   the total rms magnetic field, $\brac{B}$ the mean magnetic
   field, and $\brac{b_xb_y}$ are the turbulent stresses with the
   contribution of the mean fields removed.}
\label{nonhel64neg}
\begin{tabular}{llcccccc}\hline
$\kef/k_1$ &$f_0$ &$\brac{u_xu_y}$ &$\mu_0^{-1}\brac{b_xb_y}$ &$-M_{xy}/R_{xy}$ &$u_{\rm rms}/c_{\rm s}$ &$B_{\rm rms}/(\sqrt{\mu_0 \rho_0} c_{\rm s})$ &$\brac{B}/B_{\rm rms}$\\ \hline
1.5 &0.01   &-2.04 (-5) &-5.49 (-5) &-2.70 &0.10 &0.11 &0.63 \\
    &0.025  &-3.77 (-5) &-3.29 (-4) &-8.72 &0.14 &0.07 &0.34 \\
    &0.05   & 2.67 (-4) &-6.88 (-4) & 2.58 &0.20 &0.08 &0.20 \\
    &0.10   & 1.28 (-3) &-1.55 (-3) & 1.20 &0.32 &0.13 &0.18 \\
    &0.15   & 2.70 (-3) &-2.51 (-3) & 0.93 &0.42 &0.17 &0.16 \\
    &0.20   & 3.73 (-3) &-3.71 (-3) & 0.99 &0.50 &0.21 &0.16 \\
    &0.25   & 6.90 (-3) &-4.03 (-3) & 0.58 &0.61 &0.23 &0.17 \\
    &0.27   & 8.17 (-3) &-4.28 (-3) & 0.52 &0.64 &0.25 &0.17\\
\hline
3 &0.01   &-8.91 (-5) &-3.47 (-4) &-3.90  &0.07 &0.09 &0.25\\ %1.5nohelfor0.01neg
  &0.025  &-1.87 (-4) &-6.74 (-4) &-3.60  &0.13 &0.09 &0.25\\ %1.5nohelfor0.025neg
  &0.05   &1.30 (-4) &-1.49 (-3)  &11.46  &0.20 &0.12 &0.17\\ %1.5nohelfor0.05neg
  &0.10   &2.10 (-3) &-2.11 (-3)  &1.00   &0.33 &0.15 &0.15\\ %1.5nohelfor0.10neg
  &0.15   &4.17 (-3) &-2.98 (-3)  &0.71   &0.43 &0.20 &0.14\\ %1.5nohelfor0.15neg
  &0.20   &6.59 (-3) &-3.61 (-3)  &0.55   &0.51 &0.23 &0.13 \\%1.5nohelfor0.20neg
  &0.25   &1.03 (-2) &-3.62 (-3)  &0.35   &0.61 &0.25 &0.15\\ %1.5nohelfor0.25neg
\hline
5 &0.01   &-1.13 (-4) &-3.07 (-4) &-2.71 &0.08  &0.11 &0.41\\
&0.025  &-1.68 (-4) &-8.46 (-4) &-5.04 &0.12  &0.11 &0.42\\
&0.05   &4.73 (-4)  &-1.26 (-3) &2.66  &0.20  &0.12 &0.29\\
&0.10   &2.82 (-3)  &-1.70 (-3) &0.61  &0.32  &0.15 &0.16\\
&0.15   &5.28 (-3)  &-2.06 (-3) &0.39  &0.42  &0.19 &0.14\\
&0.20   &7.31 (-3)  &-2.37 (-3) &0.32  &0.50  &0.22 &0.13\\
\hline
10 &0.01   &-4.34 (-5) &-1.18 (-4) &-2.73  &0.08 &0.14 &0.33\\ %1.5nohelfor0.01neg
&0.025  &-5.07 (-5) &-4.01 (-4) &-7.91  &0.10 &0.16 &0.57\\ %1.5nohelfor0.025neg
&0.05   &1.56 (-4) &-7.02 (-4)  &4.51   &0.15 &0.16 &0.67\\ %1.5nohelfor0.05neg
&0.10   &2.02 (-3) &-8.43 (-4)  &0.42   &0.28 &0.13 &0.28\\ %1.5nohelfor0.10neg
&0.15   &3.82 (-3) &-8.07 (-4)  &0.21   &0.39 &0.14 &0.13\\ %1.5nohelfor0.15neg
\hline
\end{tabular}
\end{table*}

\subsubsection{MRI-stable runs}\label{MRIstable}

We attempt to separate the stresses generated by the external forcing
from the MRI-generated ones by repeating the calculations with
MRI-stable shear profile for which the angular velocity is increasing
outwards (the sign of the $q$-parameter is reversed, so that
$q=-3/2$). 
All the other parameters are kept unchanged; the results from this set
of calculations are summarised in Table~\ref{nonhel64neg}.

As in the MRI-unstable calculations, the MRI-stable calculations show
large-scale dynamo action due to which a mean magnetic field is generated. The
negative-$q$ dynamo appears to be more efficient than the positive-$q$ one; the
strength of the mean magnetic field reaches 60 percent of the total
magnetic field strength in some cases. To calculate the turbulent
stresses, we have removed the contribution of the mean field using
Eq.~(\ref{removemf}).

\begin{figure}
\includegraphics[width=80mm,height=50mm]{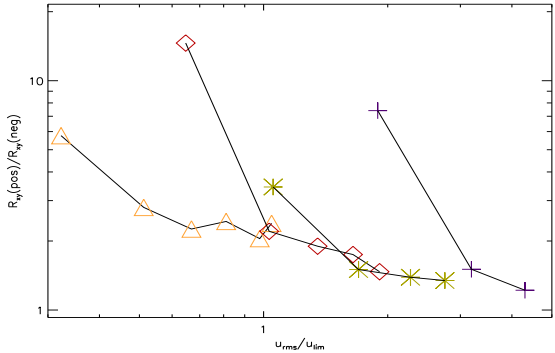}
\caption{Reynolds stresses from the MRI-unstable runs normalised by stresses
  generated in the MRI-stable runs with corresponding forcing
  wavenumber and amplitude. Colors and symbols as in Fig.~\ref{rey_pure}.}
\label{rey_neg}
\end{figure}

\begin{figure}
\includegraphics[width=80mm,height=50mm]{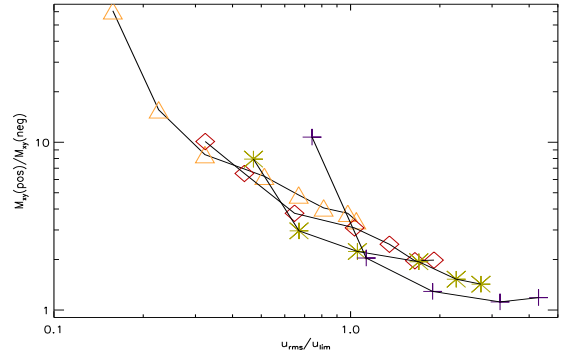}
\caption{Maxwell stresses from the MRI-unstable runs normalised by stresses
  generated in the MRI-stable runs with corresponding forcing
  wavenumber and amplitude. Colors and symbols as in Fig.~\ref{rey_pure}.}
\label{max_neg}
\end{figure}

For the weakest forcings, the Reynolds stress is small and negative,
showing similarity to the hydrodynamic positive $q$ runs presented in
Sect.~\ref{hydro}. For stronger forcings, positive values,
monotonically increasing with the forcing amplitude, are obtained. The
Maxwell stress is always negative, and monotonically increasing in
magnitude, but beginning to show sings of saturation for the largest
forcing amplitudes. We use these stresses to normalise the ones of
the MRI-unstable runs; the results are plotted in Figs.~\ref{rey_neg}
and \ref{max_neg}. The trend is quite clear for both of the stresses:
for the weakest levels of turbulence, there is an excess both of
Reynolds and Maxwell stresses in the MRI-unstable versus the
MRI-stable case, indicative of ongoing MRI-activity. As the forcing is
increased, especially the Reynolds stress quickly levels off to a
roughly constant value, not showing clear correlation with the $u_{\rm
  lim}$. The saturation level of the Reynolds stress is close to one
only in the set with the largest wavenumber $\kef/k_1=10$, and
increasing with decreasing $\kef$. This can be attributed to the
asymmetry $q \rightarrow -q$ observed also in the hydrodynamic
calculations (Sect.~\ref{hydro}). The Maxwell stress also shows a
monotonically decreasing trend, the level-off occurring roughly at
$u_{\rm lim}$. The saturation level, again, is close to unity only for
the set $\kef/k_1=10$, and is increasing with decreasing
wavenumber. We relate this behavior to the $q \rightarrow -q$
asymmetry; due to this asymmetry, no definite conclusion about the
damping limit of the MRI can be drawn even with the help of these
reference calculations.

\section{Application to the Milky Way}\label{MW}

In addition to accretion disks, the disks of spiral galaxies are also
subject to the MRI, as the angular velocity is decreasing outwards so
that the rotational velocity is roughly constant as a function of
radius outside the very centre (called as flat rotation curve, for
which $q=1$). Sellwood \& Balbus (1999) proposed that the MRI could be
responsible of the anomalous velocity dispersions observed in the
outer regions of some galaxies, and later, in a series of papers,
Piontek \& Ostriker showed that the MRI can really work in the cool
cloudy and clumpy medium subject to thermal instability (see
e.g. Pion\-tek \& Ostriker 2007). The energy balance estimates from the
observations of NGC6949 (Beck 2004) indicate that magnetic energy
could become dominant over the kinetic energy in the outer regions of
this galaxy, so that the energy ratio $E_{\rm M}/E_{\rm K} \approx
3$--$4$.  This matches well the energy ratios found from numerical
simulations of MRI (e.g.\ Liljestr\"om et al.\ 2009).

In the inner star-forming regions of spiral galaxies there is another
very powerful source of turbulence in addition to the MRI, namely
supernova explosions (hereafter SN). SN-forced turbulent flows have
been studied numerically (e.g.\ Korpi et al.\ 1999; de Avillez \& Mac
Low (2002); Gressel et al.\ 2008), but no signs of MRI has been
reported yet. Here we apply the linear analysis result together with
the FOSA-prediction to obtain a preliminary estimate whether the MRI
could be suppressed by SN activity.

\begin{figure}
\includegraphics[width=80mm,height=50mm]{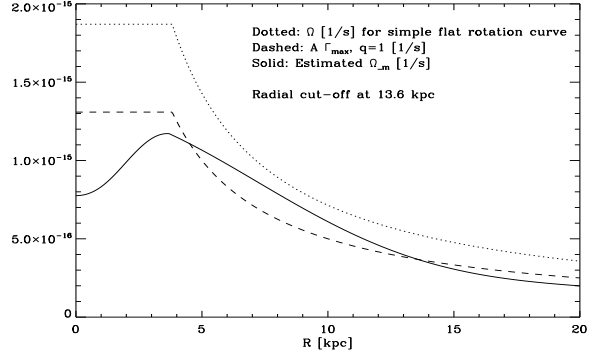}
\caption{Galactic application.
An approximate fit to the galactic rotation profile (dotted line), the
resulting $\Omega_{\rm m}$ from Eq.~(\ref{OmFOSA}) (solid line), and
the damping curve $A\ \Gamma_{\rm max}=A/2\ \Omega(R)$ (see
Eq.\ref{dampcond}) (dashed line).}
\label{galaxy}
\end{figure}

For such an analysis, we need the rotation law for the determination of
 $\Omega$ and the expected growth rate of the MRI, the SN distribution as a
function of radius, and the expected rms-velocity and scale of SN driven 
turbulence as functions of the SN forcing. For the
rotation law we choose a simplified fit to the galactic rotation
profile similar to the one used by Ferri\`ere \& Schmitt (2000), shown
in Fig.~\ref{galaxy}. 
The radial distribution of SNe we obtain from the review article
of Ferri\`ere (2001), who gives
\begin{eqnarray}
\sigma_{\rm I}(R) &=& \left( 4.8\,{\rm kpc}^{-2}{\rm Myr}^{-1} \right)
\nonumber \label{eq:snI}\\
&&\times {\rm exp} \left( -\frac{R-R_{\odot}}{4.5\,{\rm kpc}} \right)\\ 
\sigma_{\rm II} (R<3.7\,{\rm kpc}) &=& \left( 27\,{\rm kpc}^{-2}{\rm Myr}^{-1}\right) \nonumber \\
&&\times 3.55 \ {\rm exp} \left( -\frac{R-3.7\,{\rm kpc}}{2.1\,{\rm kpc}}\right)^2 \\
\sigma_{\rm II} (R>3.7\,{\rm kpc}) &=& \left( 27\,{\rm kpc}^{-2}{\rm Myr}^{-1}\right) \nonumber \\
&&\times {\rm exp} \left( -\frac{R^2-R^2_{\odot}}{(6.8\,{\rm kpc})^2}\right),\label{eq:snII}
\end{eqnarray}
where $R_{\odot}=8.5$~kpc is the radius of the solar orbit.

Dib et al.\ (2006) investigated the velocity dispersion resulting from
SNe of varying frequency in a cube of nonstratified and nonmagnetized
gas. No rotation or differential rotation was included, so the system
can be regarded as isotropic. SN frequencies of 0.01 to 10 times the
frequency at the solar radius ($R_{\odot}=8.5$~kpc) were investigated
and velocity dispersions, but no information of the scale of the
turbulence, were reported. From their plots, however, it is evident
that the scale is approximately 100 pc with the SN frequency at solar
neighborhood.
Gressel (2009) also arrives at a similar value by computing velocity
structure functions from more sophisticated simulations of SN forced
stratified flows.
For our simple analysis, we therefore adopt the value
$l_{\rm f}=100$~pc, but neglect any variation as a function of SN
frequency. We then perform an approximate fit to the velocity
dispersion as a function of SN frequency from the Figure 13 of Dib et
al.\ (2006), and map the used SN frequencies over the galactic radius
using the SN-distributions Eqs.~(\ref{eq:snI}) -- (\ref{eq:snII}). The
resulting $\Omega_{\rm m}$ from Eq.~(\ref{OmFOSA}) is plotted with a solid
line in Fig.~\ref{galaxy}.

Finally, from the linear stability analysis, we choose $A=1.4$,
corresponding to the limit where we found all the MRI-modes to be
damped for $q=1$ (see Fig.~\ref{LSA_damp}). We show the corresponding
damping curve $A\ \Gamma_{\rm max}=A/2\ \Omega(R)$ (see
Eq.~(\ref{dampcond})) in Fig.~\ref{galaxy} with a dashed line. MRI is
predicted to be damped by SNe in the regions where the solid line is above
the dashed line. This occurs up to the radius of roughly 13.6~kpc.

The analysis presented above can be improved by including a more
realistic description of the turbulent scale and rms-velocity and
their dependences, e.g.\, on the interstellar magnetic field strength
and surrounding density, aspects that were totally neglected
here. This would require high-resolution numerical simulations of SN
forced flows, involving a huge parameter space to explore. Intuitively
one would expect, however, that the presence of a large-scale magnetic
field and larger surrounding density would probably both result in
smaller values of rms-velocities and sizes of the individual SN
shells. Therefore the ratio $u_{\rm rms}/l_{\rm f}$, appearing in the
Ohmic diffusion rate in the isotropic case (Eq.~(\ref{OmFOSA})), 
could remain unchanged.

\section{Conclusions} \label{conc}

In this study we have investigated the effect of turbulent Ohmic
diffusion on the excitation and dynamics of the MRI. First we
presented a linear stability analysis for different types of rotation
profiles of the form $\Omega \propto R^{-q}$, and proceeded by making
a prediction in the isotropic homogeneous case of the level of
turbulence needed to suppress MRI (Eq.~(\ref{LSApred})). 

Next we made an attempt, with the help of three-dimensional numerical
simulations, to investigate whether this simple criterion holds for
more complex nonlinear systems. We began with purely hydrodynamic
calculations with both positive (angular velocity decreasing outwards)
and negative (angular velocity increasing outwards) shear parameters
$q$. These calculations gave two important results: Firstly, non-zero
Reynolds stresses arise solely due to the interaction of forcing,
shear and rotation, in agreement with the earlier results of Snellman
et al.\ (2009), but in apparent disagreement with the study of Workman
\& Armitage\ (2008). Secondly, there is a strongly $\kef$-dependent
asymmetry in the generated stresses in a transition $q \rightarrow
-q$, so that the smaller the effective forcing wavenumber, the larger
the asymmetry is in the generated Reynold stress. In the MHD regime,
therefore, non-zero stresses can occur both from the forcing and the
MRI, and the separation of the two contribution is very difficult, as
no absolute reference point can be established due to the $q
\rightarrow -q$ asymmetry.

In the magnetohydrodynamical regime sets of MRI-unstable (positive
$q$) runs were made, and compared to a calculation without external
forcing (exhibiting only MRI). As anticipated, the interpretation of
the results without any other reference point than the pure MRI case,
was difficult. Utilising the corresponding hydrodynamic calculations
as reference, the MRI versus the forcing contributions to the Reynolds
stress could be investigated in more reliable manner. From this
analysis, indications of MRI damping were found. For the Maxwell
stress, a normalisation to the rms magnetic field strength of each
calculation was found to be a useful diagnostic, giving further
support to the damping scenario. The Maxwell to Reynolds stress ratio
was also observed to be a sensitive diagnostic of the MRI
efficiency. Negative-$q$ calculations were also made in the MHD
regime; due to the $q \rightarrow -q$-asymmetry of the generated
stresses, these runs, however, serve as a rather poor reference
point. 
It is possible to study this issue further with the help of nonlinear
closure models, such as Ogilvie (2003) or Snellman et
al. (2009). The closure model of the latter study could be used to
investigate the $q \rightarrow -q$-asymmetry in more detail, whilst
the former could serve as a tool to separate the two stress
contributions; this, however, is out of the scope of this paper.

All in all, the effect of the external forcing on the excitation and
dynamics of the MRI is to reduce its efficiency as angular momentum
transport mechanism (seen both in the Reynolds and Maxwell stresses,
and most clearly in their ratio), and as the turbulence level exceeds
$u_{\rm lim}$, derived by linear stability analysis and the
first-order smoothing approximation, to suppress the MRI-activity. To
determine the exact level of turbulence at which the MRI becomes damped
in the three-dimensional calculations is not possible due to the $q
\rightarrow -q$-asymmetry, but according to our results, the
LSA-FOSA-prediction is quite a reliable indicator. 

As evident, the conclusions of our study are in complete disagreement
with the very similar study of Workman \& Armitage\ (2008). The
discrepancy is mostly due to the results of the hydrodynamical
calculations of both studies: our calculations clearly show non-zero
$R_{xy}^{\rm f}$, whereas Workman \& Armitage\ (2008) find very small
$R_{xy}^{\rm f}$, which is interpreted to be consistent with
zero. 
This likely to be explained by the fact that their conclusion is based
on only one point although the stress varies non-monotonically as a 
function of $\Omega$ and $S$ (Snellman et al.\ 2009). 
Although the results of the magnetohydrodynamical
calculations of both studies are in perfect agreement, the conclusions
become severely altered due to the different reference points. 

One particularly interesting case to apply our result is that of the
star-forming inner parts of galactic disks, where the stellar
turbulent energy input produces turbulence via stellar winds,
radiation pressure and supernova explosions. On the other hand, the
galactic disks are susceptible to the MRI with flat rotation profiles
of the type $q=1$; is the stellar energy input enough to suppress the
MRI, and over which range of radius does this occur? In this study we
made a crude estimation utilising the LSA-FOSA-prediction together
with earlier results from numerical investigations of SN forced
turbulence (Korpi et al.\ (1999); Dib et al.\ (2006)) and the radial
SN distribution reported in the litterature (Ferri\`ere 2001). Our
analysis gives indication that supernova activity can suppress the MRI
up to the radius of roughly 14 kpc. 

\acknowledgements{The simulations were performed with the computers
  hosted by CSC, the Finnish IT center for science financed by the
  Ministry of Education. Financial support from the Academy of Finland
  grants No.\ 121431 (PJK) and 112020 (MJK) is acknowledged. We thank
  Oliver Gressel and Jared Workman for helpful comments on the
  manuscript.}

\end{document}